# Quantitative Complementarity of Wave-Particle Duality

Tai Hyun Yoon[1,2,*] and Minhaeng Cho[1,3,†]

[1]Center for Molecular Spectroscopy and Dynamics, Institute for Basic Science (IBS), Seoul 02841, Republic of Korea

[2]Department of Physics, Korea University, Seoul 02841, Republic of Korea

[3]Department of Chemistry, Korea University, Seoul 02841, Republic of Korea

To test the principle of complementarity and wave-particle duality quantitatively, we need a quantum composite system that can be controlled by experimental parameters. Here, we demonstrate that a double-path interferometer consisting of two parametric downconversion crystals seeded by coherent idler fields, where the generated coherent signal photons are used for quantum interference and the conjugate idler fields are used for which-path detectors with controllable fidelity, is useful for elucidating the quantitative complementarity. We show that the source purity $\mu_s$ is tightly bounded by the entanglement measure $E$ by the relation $\mu_s = \sqrt{1-E^2}$ and the visibility $V$ and detector fidelity $F$ determine the coherence of the quantons, i.e., $C = V|F|$. The quantitative complementarity of the double-path interferometer we developed recently is explained in terms of the quanton-detector entanglement or quanton source purity that are expressed as functions of injected seed photon numbers. We further suggest that the experimental scheme utilizing two stimulated parametric downconversion processes is an ideal tool for investigating and understanding wave-particle duality and complementarity quantitatively.

**Introduction**

Bohr's principle of complementarity [1] was initially a qualitative statement about mutually exclusive but equally real properties of a single quantum object, the quanton, such as photons, electrons, etc. [2]. Later in 1979, Wooters and Zurek proposed the wave-particle duality concept [3], as a quantitative version, i.e., experimentally testable complementarity relation, in its best-known representative. After that, by considering an unbalanced two-beam interferometer [4], where two beams have not equal intensities, the concept of path predictability $P$ is introduced for about which-path an interfering quantum object takes. This predictability limits the amount of visibility $V$ that can be achieved in an interferometer according to the complementarity inequality relation, $P^2+V^2 \leq 1$. This inequality becomes an equality if the quantum object can be described as a pure state, which is the case that the wave-particle duality becomes a quantitative statement. This relation has been generalized to the systems where which-way detectors are in place. Then, another property called path distinguishability $D$ takes the role of path predictability $P$ to represent particle nature of the quanton as detailed in double-path interferometers [5-8], multi-path interferometers [9-11], and delayed choice quantum erasing schemes [12-17].

Recently, systematic approaches to establishing *quantitative* complementarity relations in various composite quantum systems have been reported [8,18-21]. In generalized complementarity relations, multi-partite realities mutually exclude single-partite properties of the subsystems, and the complementarity was shown to be quantified by entanglement measures between subsystems [8]. For example, the entanglement measure appears in the form of the concurrence for a bipartite system, which becomes an essential entry in the quantitative complementarity relation consisting of path predictability $P$, quantum coherence $C$, and entanglement $E$ [19]. Here, the $E$ can be a measure of entanglement between the quanton and its internal states [22] or polarization states [23] or spin states [24], or which-path detector states [11]. Although such quantitative complementarity relations proposed recently require an experimental system enabling one to measure not only the entanglement measure [7,8,9] but also the quantum coherence reflecting interference contrast or visibility [25,26] of a specific composite system, a lack of composite system whose quantum states can be experimentally controlled makes it challenging to characterize wave-particle duality involving the well-known interference fringe visibility $V$ and path distinguishability $D$. More specifically, an analytical model of the superposition state of the quantons as a subsystem of the composite system with well-defined which-path detectors or source impurity states is still missing for complete experimental verification of quantitative complementarity.

Here, we show that our quantum optical system (see Figure 1 for a schematic representation of the setup) that was used to demonstrate frequency-comb single-photon interferometry before is suitable for studying and confirming the *quantitative* complementarity relations reported recently [11,27]. We used a pair of identical but otherwise independent *nonlocal* quantum sources, i.e., parametric down-conversion (PDC) crystals, that are pumped by synchronized optical frequency-comb lasers with the same amplitudes. The PDC processes of the two nonlinear crystals were simultaneously stimulated by highly coherent lights with different but controllable amplitudes, which were generated from a single coherent laser with an extremely narrow bandwidth (see Supplementary Information for experimental details). Using this pair of stimulated PDC quantum sources, we could generate a coherence-tunable superposition state of signal single-photons and conjugate single-photon-added coherent states (SPACS) in an idler mode, where the idler photon states turned out to play the role of which-path detectors with adjustable fidelity, *F*, or impurity states degrading the source purity of quantons, i.e., signal photons. We further prove that the source purity, denoted as $\mu_s$, introduced by Qian and Agarwal [27] is related to the entanglement measure *E* representing the quantum correlation between the quanton and path-detectors as $\mu_s = \sqrt{1 - E^2}$. Also, the coherence *C* of quantons, i.e., signal photons, is connected to the well-known interference fringe visibility as *C* = |*F*|*V*, where the proportionality constant is the fidelity determined by the overlap of the entangled which-path detector states or impurity states, i.e., idler photon states, that are not necessarily orthogonal to each other.

**Double-path interferometer with single-photon source and entangled path-detectors**

The composite quantum system that will be referred to as the entangled nonlinear bi-photon source (ENBS) model is shown in Figure 1. The ENBS consists of two spatially separated type-0 phase-matched spontaneous PDC (SPDC) crystals (Figure 1), i.e., periodically-poled lithium niobate crystals (PPLNs) [28], that are the entangled signal single-photon generation sources as well as the conjugate idler photons acting like which-path detectors [11]. Hereafter, we shall refer to the signal photons as quantons because they are subject to a double-path interferometric detection in the ENBS (see Supplementary materials). The idler modes of two PPLNs are seeded by weak lasers in coherent states $|\alpha_1\rangle$ and $|\alpha_2\rangle$, which enables us to control the overlap of the two idler states precisely. The generated signal photons are used for double-path quantum interference experiments, whereas their conjugate idler photons in a SPACS provide the which-path information. Through adjusting the seed beam amplitudes, we could control the overlap between the SPACS and the single-photon-non-added coherent state [29,30]. The signal photons propagate in free space following precisely the same plane-wave spatial modes of the idler fields, which contrasts with the diffraction waves in typical double-slit experiments, which allows us to achieve almost perfect single-photon visibility [28]. It should also be noted that unlike the previous cases utilizing path-detectors for extracting which-path information of the quanton, the single photons do not have to interact with external devices [7] on their paths from the corresponding source (PPLN) to a single-photon detector (PD) (Figure 1).

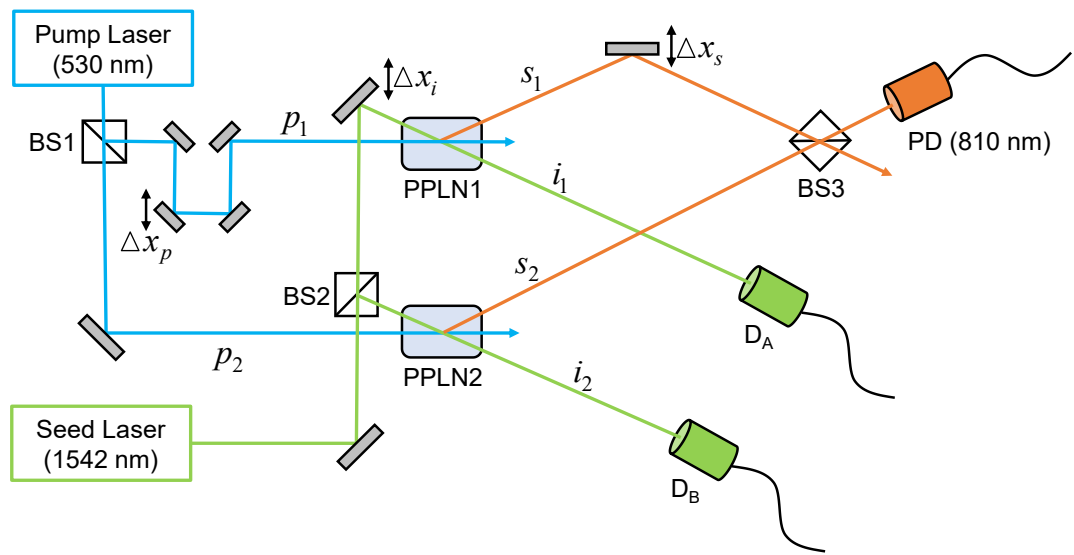

**Fig. 1. Double-path single-photon interferometer with controllable source purity used in our ENBS model.** Two SPDC crystals, PPLN$_1$ and PPLN$_2$, pumped and seeded simultaneously by the same pump and seed coherent lasers, respectively, resulting in the emission of signal photons $s_1$ or $s_2$ for quantum interference detection at detector PD. Then, conjugate idler photons $i_1$ and $i_2$ provide the which-path (or which-source) information, where the controllable source purity is determined by the overlap of the SPACS and unchanged coherent state of the idler fields. Two idler fields can be detected independently by detectors D$_A$ and D$_B$.

In our ENBS, the energy and momentum conservation relations at two PPLNs are assumed to be matched perfectly between the pump, signal, and idler fields, i.e., $\omega_p = \omega_s + \omega_i$. The seed beam frequency in an idler mode is also matched well within the emission spectrum of the idler fields. Let us consider the case that only one signal photon is generated from either

PPLN$_1$ or PPLN$_2$ at a time, which is achievable by reducing the mean photon number of stimulating coherent seed laser. The conjugate idler photon can then be considered to be a which-path detector because a pair of signal and idler photons must be simultaneously generated by annihilation of a single pump photon via an SPDC process [31]. We demonstrated that this double-path interferometer could generate a superposition state of signal single-photons. Using the following notation $|p,q\rangle_s = |p\rangle_{s_1}|q\rangle_{s_2}$ for $(p,q) \in \{0,1\}$ = {vacuum, single photon} for each single-photon state, one can write the quanton's superposition state as

$$|\psi_0\rangle = c_1(\alpha_1, \alpha_2)|1,0\rangle_s + c_2(\alpha_1, \alpha_2)|0,1\rangle_s, \tag{1}$$

where $c_j(\alpha_1, \alpha_2) = \sqrt{1 + |\alpha_j|^2}/\sqrt{2 + |\alpha_1|^2 + |\alpha_2|^2}$ and $\sum_{j=1}^{2}|c_j(\alpha_1,\alpha_2)|^2 = 1$. The probability of finding a signal single-photon from each PPLN is constant per single-photon detection at the detector PD in Figure 1 regardless of the pump and seed beam intensities after the normalization of $|\psi_0\rangle$ with the emission rate per unit time interval that strictly depends on both intensities. The signal single-photon states $|1,0\rangle_s$ and $|0,1\rangle_s$ form an orthonormal set. When the quanton emerges either one of two PPLNs pumped by the same laser with equal intensities but with different coherent seed lasers with $|\alpha_1| \neq |\alpha_2|$, the composite system of quanton and path-detector (or source impurity) can be represented by the following superposition state,

$$|\psi\rangle = c_1(\alpha_1, \alpha_2)|1,0\rangle_s|d_1\rangle + c_2(\alpha_1, \alpha_2)|0,1\rangle_s|d_2\rangle, \tag{2}$$

where the entangled idler (path-detector) states are

$$|d_1\rangle = \hat{a}_{i_1}^\dagger|\alpha_1\rangle_{i_1}|\alpha_2\rangle_{i_2} = |\alpha_1, 1\rangle_{i_1}|\alpha_2\rangle_{i_2}, \tag{3a}$$

$$|d_2\rangle = \hat{a}_{i_2}^\dagger|\alpha_1\rangle_{i_1}|\alpha_2\rangle_{i_2} = |\alpha_1\rangle_{i_1}|\alpha_2, 1\rangle_{i_2}. \tag{3b}$$

Here, $\hat{a}_{i_j}^\dagger$ is the creation operator of the $j$th idler field, and $|\alpha_j, 1\rangle_{i_j} = \frac{\hat{a}_{i_1}^\dagger|\alpha_j\rangle}{\sqrt{1+|\alpha_j|^2}}$ is the SPACS of the $j$th idler field created by the stimulated PDC of PPLN$_j$ [29]. $|d_j\rangle$ corresponds to the product idler-state, where the idler field $i_j$ is in the SPACS $|\alpha_j, 1\rangle_{i_j}$ while the other idler field $i_{k \neq j}$ is in the incident coherent state $|\alpha_{k \neq j}\rangle_{i_{k \neq j}}$. It should be emphasized that this entanglement between the quanton and which-path detector states is a fundamental requirement of the process of measurement as laid down by von Neumann [8]. We believe that our ENBS

experimentally studied before to demonstrate single-photon interferometry with optical frequency-comb pump, monochromatic seed laser-induced idler photons, and PDC-generated signal single-photons is a useful system for creating an entangled state (Eq. (2)) between the quanton and which-path detectors with controllable fidelity.

From the superposition state, we could obtain the reduced density operator of the quanton by tracing over the path-detector states

$$\rho_s = \sum_{i=1}^{2} \sum_{j=1}^{2} c_i(\alpha_1, \alpha_2)^* c_j(\alpha_1, \alpha_2) \langle d_i | d_j \rangle \rho_{ij}, \qquad (4)$$

where $\rho = |\psi_0\rangle\langle\psi_0|$ is the density operator of the pure quanton state (Eq. (1)) spanned by the basis states $\{|1,0\rangle_s, |0,1\rangle_s\}$. With the ENBS system described by Eqs. (2) and (4), we shall show that the wave-particle duality or triality can be quantitatively studied and that various relations between $E$ and $_s$ and between $C$ and $V$ can be expressed in closed forms.

**Complementarity from path-detector point of view**

Recent studies have focused on completing wave-particle duality relations using entanglement and polarization. Eberly and coworkers [23,32] showed that, in the classical optics regime, polarization was taken into consideration in the two-slit interference experiment and the triality relation among the predictability, interference visibility, and concurrence. De Zela [33] investigated the relationship between polarization indistinguishability and entanglement. More recently, for an $n$-path interference system with a path-detector that can be represented as $|\psi\rangle = \sum_{j=1}^{n} c_j |\phi_j\rangle |d_j\rangle$ where $|\phi_j\rangle$ is the state corresponding to the quanton taking the $j$th path and $\{|d_j\rangle\}$ are certain normalized states of the path-detector, Qureshi showed that the distinguishability $D$, the path predictability $P$, and the entanglement $E$ between the quanton and path-detector satisfies the Pythagorean relation $D^2 = P^2 + E^2$ [11]. Here, it should be emphasized that $\{|\phi_j\rangle\}$ for an orthonormal set of quanton states, but the detector states $\{|d_j\rangle\}$ are not necessarily orthogonal to one another. From the definition of the coherence $C$ of the composite state [26,34], generalized duality equality for the $n$-path interferometer ($n \geq 2$) was proposed as $D^2 + C^2 = 1$ or equally $P^2 + E^2 + C^2 = 1$. Although the importance of entanglement in the quantitative complementary relation was theoretically clarified in these works, the most important question about how to realize such an interferometer with well-defined detector states experimentally was not addressed.

Here, we show that our ENBS is one of the double path interferometers ($n = 2$). For the composite state $|\psi\rangle$ of the ENBS, we find that the distinguishability $D$, predictability $P$, entanglement $E$, visibility $V$, and coherence $C$, can be written, respectively, as

$$D^2 = 1 - \left(\sum_{i \neq j} \sqrt{\rho_{ii}\rho_{jj}}|\langle d_i|d_j\rangle|\right)^2 = 1 - 4\rho_{11}\rho_{22}|F|^2, \tag{5a}$$

$$P^2 = 1 - \left(\sum_{i \neq j} \sqrt{\rho_{ii}\rho_{jj}}\right)^2 = 1 - 4\rho_{11}\rho_{22}, \tag{5b}$$

$$E^2 = \left(\sum_{i \neq j} \sqrt{\rho_{ii}\rho_{jj}}\right)^2 - \left(\sum_{i \neq j} \sqrt{\rho_{ii}\rho_{jj}}|\langle d_i|d_j\rangle|\right)^2 = 4\rho_{11}\rho_{22}(1 - |F|^2), \tag{5c}$$

$$V = \sum_{i \neq j}|\rho_{ij}| = 2|\rho_{12}|, \tag{5d}$$

$$C = \sum_{i \neq j}|\rho_{ij}||\langle d_i|d_j\rangle| = V|F|. \tag{5e}$$

Here, we introduce the fidelity $F (= \langle d_i|d_j\rangle)$ of the two idler states, which is the overlap of the two path-detector states [29,30]. For the ENBS, the fidelity is the function of eigenvalues of the two coherent states describing the seed beams [28], i.e.,

$$|F| = \left|{}_{i_1}\langle\alpha_1, 1|{}_{i_2}\langle\alpha_2|\alpha_1\rangle_{i_1}|\alpha_2, 1\rangle_{i_2}\right| = \frac{|\alpha_1\alpha_2|}{\sqrt{1+|\alpha_1|^2}\sqrt{1+|\alpha_2|^2}}. \tag{6}$$

The density matrix elements of the quanton (signal photon) in the ENBS are

$$\rho_{jj} = \frac{1+|\alpha_j|^2}{2+|\alpha_1|^2+|\alpha_2|^2}, \tag{7a}$$

$$|\rho_{12}| = \frac{\sqrt{(1+|\alpha_1|^2)(1+|\alpha_2|^2)}}{2+|\alpha_1|^2+|\alpha_2|^2}. \tag{7b}$$

Inserting Eqs. (6) and (7) into (5), one can show that the five properties, $D$, $P$, $E$, $V$, and $C$ are fully determined by the two experimentally controllable eigenvalues $|\alpha_1|$ and $|\alpha_2|$ of the coherent seed beams (see Fig. 2 for the corresponding plots and Supplementary note 2 for complete expressions). From Eqs. (5), it becomes clear that the Pythagorean relation $D^2 = P^2 + E^2$ proposed by Qureshi enables one to rewrite the wave-particle duality equality $D^2 + C^2 = 1$ into the triality equality $P^2 + E^2 + C^2 = 1$ [11]. Here, it is interesting to note that the coherence $C$ representing the wave nature of the quanton of the composite state (not pure state) is related to the interference visibility $V$ of the pure state via the relation (Eq. (5e)) $C = V|F|$.

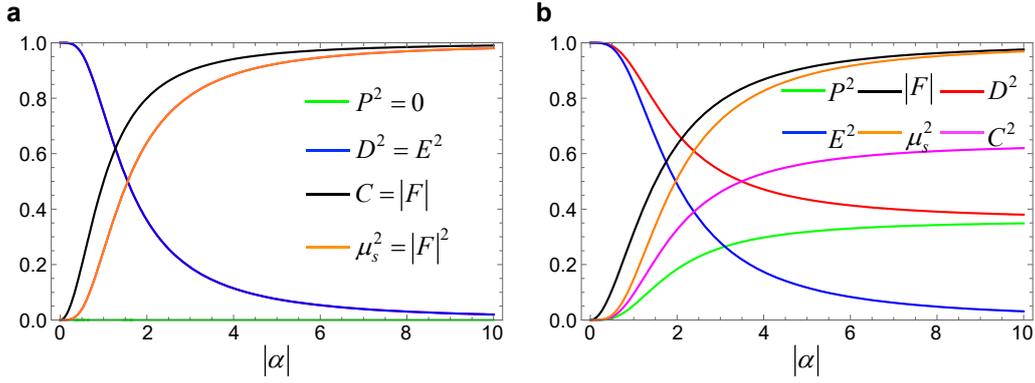

**Fig. 2. Numerical values for various parameters versus $|\alpha|$.** Parameters, $D^2$, $P^2$, $E^2$, $C^2$, $|F|$, and $\mu_s^2$, appearing in wave-particle duality relation versus $|\alpha|$, amplitude of seed beam with $\alpha_1 = \alpha_2 = |\alpha|$ (**a**) and $\alpha_1 = |\alpha| = 2\alpha_2$ (**b**). Colors associated with the measures are the same in (**a**) and (**b**).

It becomes possible to address various duality relations between distinguishability $D$, predictability $P$, coherence $C$, and entanglement $E$ quantitatively. First, let us consider the case that there is no way in principle to distinguish the signal photon emitted from which source, PPLN$_1$ or PPLN$_2$. This case corresponds to the limit that the fidelity $|F|$ equals one or to the limit of $|\alpha_1| = |\alpha_2| \gg 1$. Then, the ENBS composite system's quantum state can be written as a product of $|\psi_0\rangle$ and $|d\rangle$ ($= |d_1\rangle = |d_2\rangle$), which means that there is no entanglement between the quanton source and path-detector states. In this case, distinguishability $D$ and predictability $P$ becomes identical, i.e., $D = P$. They depend only on the probabilities $\rho_{11}$ and $\rho_{22}$ for path 1 and 2, respectively, that are constrained by the normalization condition $\rho_{11} + \rho_{22} = 1$. In addition, the coherence $C$ becomes identical to the usual visibility $V$ of the single-photon interference fringe, i.e., $C = V$.

Reducing the seed beam intensities at the two nonlinear crystals, we could experimentally control the fidelity to be in the range of $0 \le |F| < 1$. The elements in the Qureshi's Pythagorean relation $D^2 = P^2 + E^2$ can be quantitatively controlled by adjusting the overlap of the path-detector states or the path-detector fidelity $|F|$. Also, coherence $C$ and interference visibility $V$ that are two measures of wave nature of the quanton can be tuned simultaneously by varying $|\alpha_1|$ and $|\alpha_2|$ due to the relation $C = |F|V$ with Eq. (6). If $|\alpha_1| = |\alpha_2| = |\alpha|$, i.e., two seed beam intensities are identical, then $\rho_{11} = \rho_{22} = 1/2$ so that $P = 0$. In this case, one cannot predict whether the quanton will take either path 1 or 2 regardless of the magnitude of $|\alpha|$. Then, we have $D = E = \sqrt{1 - |F|^2}$, where $|F| = \frac{|\alpha|^2}{1+|\alpha|^2}$. When $|\alpha| \gg 1$, we again reach the limit $|F| \cong 1$, i.e., $D = E \cong 0$ and $C \cong 1$, indicating the path-detector loses its role to distinguish which-path the quanton takes because the coherent state $|\alpha\rangle$ and SPACS $|\alpha, 1\rangle$ overlaps perfectly. Because the quanton taking either one of the two paths is not entangled with which-path detector, i.e., $E = 0$, the quanton propagates as a perfect coherent wave with visibility $V = 1$ and coherence $C = 1$.

If $|\alpha_1| \neq |\alpha_2|$, all the measures appear to be different and play their roles in the wave-particle duality or triality relations. Evidently, we find the relations like $D^2 = P^2 + E^2$ and $P^2 + E^2 + C^2 = 1$ hold in the whole range of $|\alpha_j|$, demonstrating that all the measures can be precisely controlled by the experimental parameters $|\alpha_j|$. In summary, here, we show that the wave-particle duality (triality) equality, i.e., quantitative complementarity, can be tested with our ENBS system, where the wave-like and particle-like behaviors of the quanton (signal photon) are tunable quantities through the experimentally adjustable path-detector fidelity $|F|$ ranging from 0 to 1.

**Complementary from source point of view**

To realize single-photon interferometry experimentally, one could prepare two-level atoms with one of the two being in an excited state. Then, the spontaneously emitted photon can produce an interference fringe. Such a fluorescence single-photon interference from entangled two-level atoms was first experimentally realized by Eichmann *et al*. [35,36] with a laser beam exciting one of the two trapped $^{198}$Hg$^+$ ions. Later, Araneda *et al*. [37] carried out the same experiment with trapped $^{138}$Ba$^+$ ions. However, the observed visibility is significantly different from unity. To explain the deviation, Qian and Agarwal [27] considered the active role of remaining degrees of freedom other than the superposition state of a singly excited two two-level atoms. They introduced another interesting Pythagorean relation between wave-particle duality and source purity, i.e.,

$$P^2 + C^2 = \mu_s^2, \tag{8}$$

where the so-called source purity is defined as $\mu_s = \sqrt{2\text{Tr}[\rho_r^2] - 1}$ with $\rho_r$ being the reduced density matrix obtained by tracing over the states representing the remaining degrees of freedom. Here, note that the $D$ and $V$ in Ref. [27] should be replaced with $P$ and $C$ when the definitions of these quantities in Eqs. (5) are used for the sake of consistency. In their analysis, two entangled two-level atoms with one of them excited scatter a single photon at a time with equal probability into path 1 or 2 in the dual-path interferometer [27]. The superposition state of the pure system can be written as $|\psi_0\rangle = c_a|e_A,g_B\rangle + c_b|g_A,e_B\rangle$, where $|e\rangle$ ($|g\rangle$) is the excited (ground) state of atom $A$ or $B$. In practice, however, pure states $|e_A,g_B\rangle$ and $|g_A,e_B\rangle$ cannot be easily realized due to different states associated with all the other remaining degrees of freedom as well as due to the presence of external parties. They, thus, considered a composite state $|\psi\rangle = c_a|e_A,g_B\rangle|m\rangle + c_b|g_A,e_B\rangle|n\rangle$, where $|m\rangle$ and $|n\rangle$ represent two sets of quantum states reflecting the entangled degrees of freedom associated with the two atoms as well as even any unspecified external fields [27]. Although the state $|\psi\rangle$ is similar to any other superposition states of various composite systems discussed above, e.g., ENBS and double-path interferometer with a path-detector, the authors could not or needed not to specify the unknown quantum states $|j\rangle$ for $j \in \{m,n\}$, much like the case that the entangled which-path detector state was not specified to establish the Pythagorean relation $D^2 = P^2 + E^2$ in Ref. [11].

The relationship between wave-particle duality measures ($D$ and $V$) and source impurity given in Eq. (8) led to a new interpretation of the wave-particle duality in the sense that the source purity $\mu_s$ of the quanton can limit the totality of the complementarity relation. In other words, the source purity saturates the totality of complementarity between the wave-like and particle-like behaviors of the quanton. If we take the external states $|m\rangle$ and $|n\rangle$ as the probe reporting us the which-source or which-path information, i.e., which atom emits a quanton, much like $|d_1\rangle$ and $|d_2\rangle$ for the double-path interferometer with a which-path detector discussed above, we can use the same relations discussed above, i.e., $D^2 = P^2 + E^2$ and $P^2 + E^2 + C^2 = 1$. From these and Eq. (8), we find that the so-called source purity $\mu_s$ introduced by Qian and Agarwal is bounded by the entanglement measure $E$ as

$$\mu_s = \sqrt{1-E^2}. \qquad (9)$$

This is another interesting relationship that has not been discussed before. The entanglement between the quanton and path-detector states can play a role in degrading the purity of the source (signal single-photon) state. In the case of the ENBS, the idler state represents the which-path detector state, and the source purity can be written as

$$\mu_s = \sqrt{1-4\rho_{11}\rho_{22}(1-|F|^2)}. \qquad (10)$$

When the two seed beam intensities are the same, i.e., $|\alpha| = |\alpha_1| = |\alpha_2|$, the source purity becomes identical to fidelity, $\mu_s = |F|$, which means that the upper limit of wave-particle duality equality is limited by the source purity or detector fidelity in the case of the ENBS.

**Coherence versus visibility**

We now briefly describe our previous experiments [28] that show a single-photon interference with nearly perfect visibility, i.e., $V = 1$, or reduced visibility $0 < V < 1$, depending on the fidelity $|F|$ of the conjugate idler photons. We didn't provide a quantitative interpretation of our experimental results in connection to the quantitative complementarity discussed in this paper. To characterize the signal photon quantum interference in terms of the idler field states, we can adapt the approach in the Heisenberg picture as shown in Refs. [38,39]. The positive-frequency part of the signal electric field operator generated in crystals PPLN$_1$ and PPLN$_2$, and incident on detector DA via the symmetric beam splitter BS3, can be written as [39]

$$E_{DA}^{(+)} = iv_1\hat{s}_1 e^{i\phi_s} + v_2\hat{s}_2 = iv_1\hat{i}_1^\dagger e^{i\phi_s} + v_2\hat{i}_2^\dagger,$$

$$E_{DA}^{(-)} = -iv_1^*\hat{s}_1^\dagger e^{-i\phi_s} + v_2^*\hat{s}_2^\dagger = -iv_1^*\hat{i}_1 e^{-i\phi_s} + v_2^*\hat{i}_2, \tag{11}$$

where $\hat{s}_j$ is the annihilation operator for the signal photons emitted from PPLN$_j$, $\hat{i}_j^\dagger$ is the creation operator for the conjugate idler photons, and constant $v_j$ is proportional to the nonlinear susceptibility of PPLN$_j$ as well as the pump field incident on the PPLN$_j$. The factor $ie^{i\phi_s}$ in Eq. (11) accounts for the relative phase change in the propagation of the signal field to the detector via reflecting off the mirror, i.e., $\phi_s = k_s \Delta x_s$.

Treating the detector as a perfectly efficient, fast enough, and broadband photo-detector, we take the signal photon count rate ($g^{(1)}(0)$, i.e., second-order field correlation) $R_{DA} = \langle E_{DA}^{(-)} E_{DA}^{(+)} \rangle$ at the detector to be proportional to the normal-ordered expectation value [39,40]

$$R_{DA} = |v|^2 [\langle \hat{i}_1 \hat{i}_1^\dagger \rangle + \langle \hat{i}_2 \hat{i}_2^\dagger \rangle + ie^{i(\phi_p+\phi_s)}\langle \hat{i}_2 \hat{i}_1^\dagger \rangle - ie^{-i(\phi_p+\phi_s)}\langle \hat{i}_1 \hat{i}_2^\dagger \rangle], \tag{12}$$

where for simplicity we take $v_1 = |v|e^{i\phi_p}$ and $v_2 = |v|$, consistent with the equal powers of the pump fields incident on the PPLN$_1$ and PPLN$_2$ but with a phase change $\phi_p$ accrued by the mirror (see Figure 1), i.e., $\phi_p = k_p \Delta x_p$. As shown in Refs. [28,38,39], we assume that the downconversion efficiency is small so that the higher-order terms except for the lowest order terms in $v$ can be ignored. Note that, even though $R_{DA}$ is the normal-ordered correlation function $\langle E_{DA}^{(-)} E_{DA}^{(+)} \rangle$, it depends on the anti-normal-ordered idler-mode operators (Eq. (12)). Therefore, the positive- (negative-) frequency part of the signal electric field operator depends on the negative- (positive-) frequency part of the idler field operator.

To account for the signal single-photon interference, we use $\hat{\iota}_1 = \hat{\iota}_{10} + \alpha_1 e^{i\phi_i}$ and $\hat{\iota}_2 = \hat{\iota}_{20} + \alpha_2$ in Eq. (12), where $\hat{\iota}_{10}$ and $\hat{\iota}_{20}$ are annihilation operators for the vacuum idler modes incident on the two PPLNs [39]. Here, $\alpha_j$ is the complex amplitude describing the seed laser fields injected into the PPLN$_j$ with the phase factor $e^{i\phi_j}$ due to the presence of mirror, i.e., $\phi_j = k_j \Delta x_j$. Note that there is no induced coherence without induced emission between the generated signal fields [40-43] since the vacuum fields at two crystals are associated with distinct modes in this parallel arrangement of two PPLNs. Therefore, $\langle \hat{\iota}_{10} \hat{\iota}_{20}^\dagger \rangle = \langle \hat{\iota}_{20} \hat{\iota}_{10}^\dagger \rangle = 0$, and from Eq. (12), we have

$$R_{DA} = |v|^2 \langle \hat{\iota}_{10} \hat{\iota}_{10}^\dagger \rangle + \langle \hat{\iota}_{20} \hat{\iota}_{20}^\dagger \rangle + |\alpha_1|^2 + |\alpha_2|^2 + i e^{i(\phi_p + \phi_i + \phi_s)} |\alpha_1||\alpha_2|$$
$$- i e^{-i(\phi_p + \phi_i + \phi_s)} |\alpha_1||\alpha_2|$$
$$= |v|^2 [2 + |\alpha_1|^2 + |\alpha_2|^2 - 2|\alpha_1||\alpha_2| \sin \Delta\theta], \tag{13}$$

where we used the boson commutation relation for the idler fields, i.e., $[\hat{\iota}_i, \hat{\iota}_j^\dagger] = \delta_{ij}$, and $\Delta\theta = \phi_p + \phi_i + \phi_s$. It is interesting to note that in our ENBS system, we can measure the coherence $C$ defined in Eq. (5) directly, not the visibility $V$, from the double-path interferometer for arbitrary values of $|\alpha_1|$ and $|\alpha_2|$. More precisely, quantum coherence $C$ [26,44], which is defined as $C = \frac{R_{DA}^{max} - R_{DA}^{min}}{R_{DA}^{max} + R_{DA}^{min}}$, can be measured experimentally as

$$C = \frac{2|\alpha_1||\alpha_2|}{2 + |\alpha_1|^2 + |\alpha_2|^2} = V|F|, \tag{14}$$

where $R_{DA}^{max}$ ($R_{DA}^{min}$) is the maximum (minimum) of the interference fringe when $\Delta\theta$ is varied within the single-photon interferometer in Figure 1. $V$ and $|F|$ were given in Eqs. (5) and (6), respectively. Equation (14) tells us that the quantum coherence $C$ can be controlled experimentally by adjusting $|\alpha_1|$ and $|\alpha_2|$. The novel relation $C = V|F|$ in Eq. (14) has not been pointed out before in any literature as far as the authors are aware of. In the special case when $|\alpha_1| = |\alpha_2| = |\alpha|$, the coherence $C$ becomes identical to the fidelity $|F|$, i.e., $C = |F| = \frac{|\alpha|^2}{1+|\alpha|^2}$. In this case, even though the coherence can vary from 0 to 1, the visibility becomes unity always, i.e., $V = 2|\rho_{12}| = 1$, regardless of the magnitude $|\alpha|$. Based on the analysis in this paper, Figure 3 in Ref. [28] turned out to be the first experimental measurement of coherence $C$ versus $|\alpha|^2$ instead of visibility $V$ versus $|\alpha|^2$. This particular set of experiments is analogous to the case when entangled two-level atoms [27] emit single photons with equal probability but with controllable fidelity $|F|$ or source purity $\mu_s = |F| = C$. Compared to the system studied in Ref. [28], our ENBS system allows us, in principle, to cover the entire parameter ranges not only of the fidelity $|F|$ but also of the coherence $C$, i.e., $0 \leq \{|F|, C\} \leq 1$. Therefore, to demonstrate the quantitative complementarity established in this paper $P^2 + C^2 = \mu_s^2 = 1 - E^2$ experimentally, one needs to measure the coherence $C$ from the single-photon interference of Eq. (13) across the two-dimensional parameter space of $|\alpha_1|$ and $|\alpha_2|$.

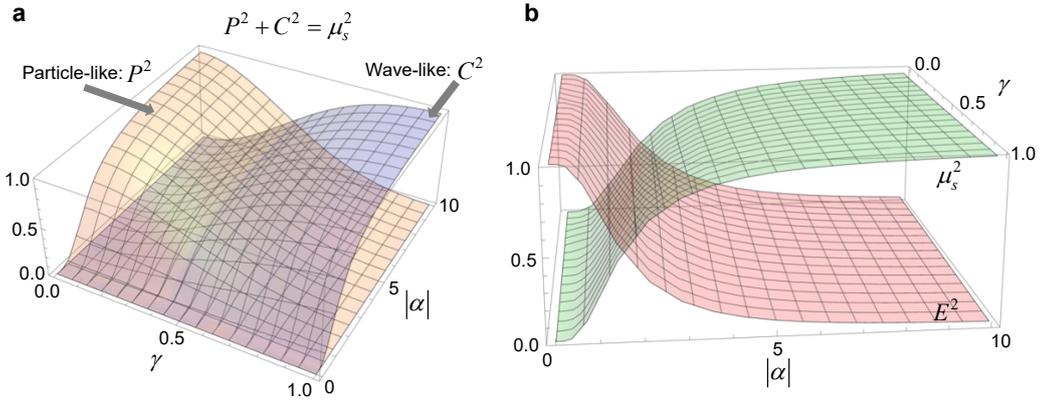

**Figure 3.** (a) Quantitative complementarity relation $P^2 + C^2 = \mu_s^2$ with respect to $\gamma = \frac{|\alpha_2|}{|\alpha_1|}$ and $|\alpha| = |\alpha_2|$. Here, path predictability $P$ represents particle-like behavior while coherence $C$ represents wave-like behavior of the quanton in the double-path interferometer. The totality of complementarity is bounded by the source purity $\mu_s$. (b) Source purity $\mu_s$ of the quanton (signal photon) and entanglement $E$ between the quanton and which-path (which-source) detector form another complementarity relation $\mu_s^2 + E^2 = 1$ These two measures are plotted with respect to $\gamma = \frac{|\alpha_2|}{|\alpha_1|}$ and $|\alpha| = |\alpha_2|$.

Indeed, Figure 4 shows the coherence $C$ in Eq. (14) in the two-dimensional space of two experimental parameters of $\gamma = \frac{|\alpha_2|}{|\alpha_1|}$ and $|\alpha| = |\alpha_2|$, where the blue symbols are the experimental data taken from Ref. [28] (In Ref. [28], we interpreted $C$ as $V$ without awareness of the roles of the entanglement $E$ and quantum coherence $C$ in the wave-particle duality of the composite system). Also shown in Figure 4 is the visibility $V$ as a function of $|\alpha|$ and $\gamma$ for comparison. From Figs. 3a and 3b, the quantitative complementarity relation $P^2 + C^2 = \mu_s^2 = 1 - E^2$ indicates that the quantum object propagates through the double-path interferometer partly as particle-like measured by $P$ and partly as wave-like measured by $C$, where their totality is bounded by the source purity $\mu_s = \sqrt{1 - E^2}$ or equally by the entanglement.

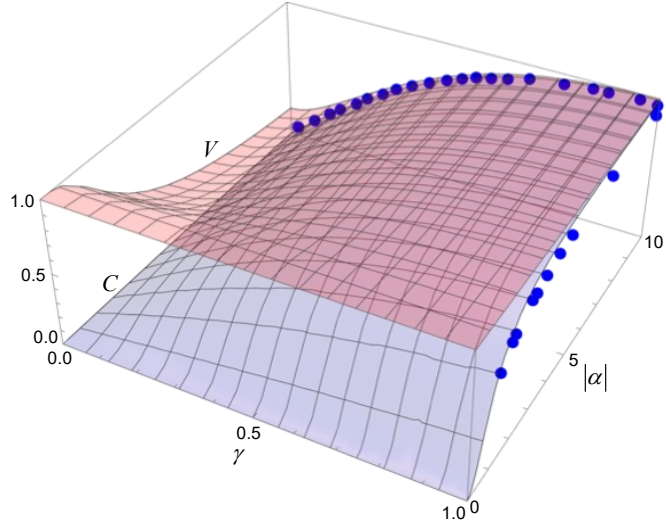

**Figure 4.** Coherence $C$ and visibility $V$ as functions of $\gamma = \frac{|\alpha_2|}{|\alpha_1|}$ and $|\alpha| = |\alpha_2|$. Blue symbols are experimental data taken from our recent paper Ref. [28] (see Figure S2). Experimental data coincide with the coherence $C$ of Eq. (14), not visibility $V$ across the whole ranges of $\gamma$ and $|\alpha|$. This plot validates our analysis of the ENBS experimental results in terms of the wave-particle duality and quantitative complementarity relations. $C$ and $V$, even though they both reflect the wave-like nature of the quanton, are different from each other in the regions of $|\alpha| < 5$ and $\gamma < 0.5$, but becomes identical when $|\alpha| \gg 1$ and $\gamma \simeq 1$.

In our ENBS system [28,38,39], the stimulated down-conversion rate can be controlled easily from the same order of spontaneous PDC to a much higher level than that of spontaneous PDC. Furthermore, the source is free from the decoherence issue because the spectrum of the quanton is determined by coherent seed beams. By adjusting the seed beam photon numbers $|\alpha_j|^2$ while fixing the pump beam intensity $|v|^2$, we have independent knobs to control the coherence $C$. These degrees of freedom in experiments enable one to enjoy additional flexibility and controllability of the quantum coherence (fidelity) of the signal photons and their emission rate with two independent (orthogonal) knobs $|\alpha_j|^2$ and $|v|^2$.

**Conclusion**

In summary, the wave-particle duality and the quantitative complementarity $P^2 + C^2 = \mu_s^2$ were analyzed and tested using our ENBS model, where the superposition state of the quanton (signal photons) is entangled with its which-path detector states (conjugate idler fields) in a controllable manner through the fidelity $|F|$. We find that the source purity $\mu_s$ depends on the entanglement $E$ between the quanton's superposition state and which-path detector states by the following relation $\mu_s = \sqrt{1 - E^2}$. We showed that path predictability $P$, quantum coherence $C$, entanglement $E$ (thus, source purity $\mu_s$, and fidelity $|F|$ in our ENBS model) depend only on the seed beam photon numbers. Furthermore, the coherence $C$ of the composite system is shown to be related to the single-photon interference fringe visibility, i.e., $C = V|F|$. We anticipate that the interpretation based on the double-path interferometry experiments with ENBS will have fundamental implications for better understanding the principle of complementarity and the wave-particle duality relation quantitatively, leading to demystifying Feynman's mystery for the double-slit experiment explanation based on the quantum mechanics [45,46].

**Materials and Methods**

A schematic diagram of the experimental setup used in Ref. [28] and in our ENBS system discussed in the main text is depicted in Fig. 5. Details of the experimental setup, not specified types of equipment, and the experimental parameters can be found elsewhere [28]. However, a brief description of the setup is presented here. Pump laser is an optical frequency comb (Pump Comb) with a center wavelength of 530 nm, repetition rate $f_{rep} = 250$ MHz, and carrier-envelop offset frequency $f_{ceo} = 20$ MHz, pulse width $\Delta\tau = 10$ ps, and optical spectral width of $\Delta\lambda = 3$ nm. Seed laser is a highly coherent continuous-wave laser at $\lambda_i = 1542$ nm and line-width of an order of Hz (manufacturer specification). Two identical PPLN nonlinear crystals (periodically-poled lithium niobate, a length of 7.9 mm, a polling period of 7.3 μm, the phase-matching temperature of 121.5 °C) are used and phase-matched for type-0 SPDC process using pump photons at $\lambda_p = 530$ nm to generate identical pairs of signal photons at $\lambda_s = 807$ nm and idler photons at 1542 nm within optical spectral widths of $\Delta\lambda_s \sim \Delta\lambda_i \sim 5$ nm. After injecting the idler seed beam ($\lambda_i = 1542$ nm) into the two crystals, both optical spectral widths $\Delta\lambda_s$ and $\Delta\lambda_i$ reduce below the instrument resolution limit of 0.1 nm. To measure the single-photon count rate at fixed pump and idler beam amplitudes and $\Delta\theta$ in Eq. (13) in the main text, the integration time $T_I$ of the photo-detector is set to be $T_I = 10$ ms. The pump amplitude $v$ and idler amplitudes $\alpha_1$ and $\alpha_2$ are adjusted to make the signal photon counts per second is less than $5 \times 10^6$, which means that at every 50 pump pulse, only one single-photon count is recorded on average. This experimental result confirms a deep single-photon generation regime. To scan the full range of coherence $C$ from 0 to 1, we needed to adjust $|\alpha_1|^2$ and $|\alpha_2|^2$ up to 100 for a fixed pump power of less than 10 mW. To determine the quantum coherence $C$ using the single-photon counting rate of Eq. (13), we vary $\Delta\theta = \phi_p + \phi_s + \phi_i$ over $2\pi$ by one of the three propagation length differences, i.e., $\Delta x_p$, $\Delta x_s$, or $\Delta x_i$ in Fig. 5, which was achieved by using one of the three different PZT (piezoelectric transducer)-mounted mirror mounts.

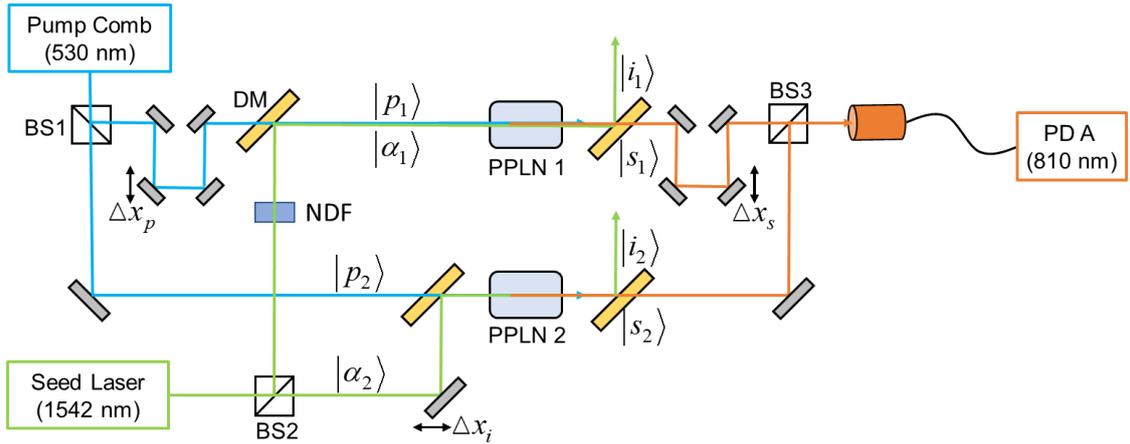

**Fig. 5. Schematic diagram of the experimental setup of the ENBS model [28].** DM; dichroic mirror, NDF; neutral density filter, BS; beam splitter, PD; single-photon sensitive photo-detector.

**Acknowledgements**

The authors acknowledge Prof. M.-S. Choi, Prof. P. Milonni, and Dr. S. K. Lee for invaluable discussion and helpful comments. This work was supported by NRF-2019R1A2C2009974 (THY) and IBS-R023-D1 (MC).